\documentclass[twocolumn,showpacs,prb,aps,epsf]{revtex4}
\usepackage{amssymb}

\usepackage{graphicx}

\begin{document}

\title{Spectral density of an interacting dot coupled indirectly to conducting leads}
\author{L. Vaugier, A.A. Aligia and A.M. Lobos}
\affiliation{Centro At\'{o}mico Bariloche and Instituto Balseiro, Comisi\'{o}n Nacional\\
de Energ\'{\i }a At\'{o}mica, 8400 Bariloche, Argentina.}

\begin{abstract}
We study the spectral density of electrons $\rho _{d\sigma }(\omega )$ in an
interacting quantum dot (QD) with a hybridization $\lambda $ to a
non-interacting QD, which in turn is coupled to a non-interacting conduction
band. The system corresponds to an impurity Anderson model in which the
conduction band has a Lorentzian density of states of width $\Delta _{2}$.
We solved the model using perturbation theory in the Coulomb repulsion $U$
(PTU) up to second order and a slave-boson mean-field approximation (SBMFA).
The PTU works surprisingly well near the exactly solvable limit $\Delta
_{2}\longrightarrow 0$. For fixed $U$ and large enough $\lambda $ or small
enough $\Delta _{2}$, the Kondo peak in $\rho _{d\sigma }(\omega )$ splits
into two peaks. 
This splitting can be understood in terms of weakly interacting
quasiparticles. 
Before the splitting takes place the universal properties of
the model in the Kondo regime are lost. Using the SBMFA, simple analytical
expressions for the occurrence of split peaks are obtained. For small or
moderate $\Delta _{2}$, the side bands of $\rho _{d\sigma }(\omega )$ have
the form of narrow resonances, that were missed in previous studies using
the numerical renormalization group. This technique also has shortcomings
for describing properly the split Kondo peaks. As the temperature is increased, 
the intensity of the split Kondo peaks decreases, but it is not completely
suppressed at high temperatures.
\end{abstract}

\pacs{72.15.Qm, 73.23.-b, 73.63.Kv}

\maketitle

\section{Introduction}

\label{Introduction} The quantum impurity models, like the Kondo and
Anderson ones have attracted the solid state physics community due to their
complex and rich behavior as well as due to their applications to many
strongly correlated physical systems.\cite{hew} In recent years, the
attention into these models has increased due to the advances in nanotechnology,
which for example made it possible to build ideal "single impurity" systems
with one quantum dot (QD) in which the Kondo physics was clearly displayed 
\cite{gold1,cro,gold2,wiel} confirming predictions based on the Anderson
model.\cite{glaz,ng,chz}. In other fascinating experiments, quantum corrals
assembled by depositing a close line of atoms or molecules on Cu or noble
metal (111) surfaces have been used to ``project" the Kondo effect to a
remote place.\cite{man} The Anderson model used to describe these systems has
the additional complication of the particular structure of the
non-interacting states, which cannot be described by a constant density of
states.\cite{rev} However, it has been shown that the non-interacting
Green's function can be written as a discrete sum of simple 
fractions.\cite{mir,hvar}

More recently, systems of several impurities or QD's have become a subject
of great interest. 
For example non-trivial results for the spectral density were
observed when three Cr atoms are placed on the (111) surface of Au.\cite
{jamn,trimer} Systems of two,\cite{jeong,craig,chen} three,\cite{waugh,gaud}
and more \cite{kouw} QD's have been assembled to study the effects of
interdot hopping on the Kondo effect, and other physical properties driven
by strong correlations. Particular systems have been proposed theoretically
as realizations of the two-channel Kondo model,\cite{oreg,zit} the so called
ionic Hubbard model,\cite{ihm} and the double exchange mechanism.\cite{mart}
In the last years, transport through arrays of a few QD's \cite
{corn,zit2,ogur,nisi,3do} and spin qubits in double QD's \cite{sq} have been
studied theoretically.

Last year, Dias da Silva \textit{et al.} \cite{dias} proposed a system with
one interacting QD (1) with a hopping term $\lambda $ to another
non-interacting QD (2), which in turn is hybridized with conducting leads.
This hybridization is in principle larger than the charging energy of QD 2,
in such a way that the latter can be considered as non interacting. The
model is equivalent to an impurity Anderson model in which the conduction
band has a Lorentzian density of states of width $\Delta _{2}$. This is one
of the simplest variations of the model in which the effects of a non
trivial structure of the conduction band can be studied. It can also be
regarded as the simplification of the Anderson model used to describe the
quantum mirage \cite{rev,mir,hvar,ali} or a tight binding model for a ring
with a quantum dot \cite{rev,pc} when only the resonance at the Fermi energy 
is included. This simplification is qualitatively
valid when the separation in energy between resonances is larger than
their width. In the case of a constant density of conduction states and on-site
energy $E_{d}$ below the Fermi energy $\epsilon _{F}$, for large enough Coulomb 
repulsion $U$, the spectral density of the impurity shows two broad peaks (called 
charge transfer peaks) at 
$E_{d} < \epsilon _{F}$ and $E_{d}+U > \epsilon _{F}$, and another peak near 
$\epsilon _{F}$ (the Kondo peak).

Previous calculations of the spectral density of the impurity in the model
for the quantum mirage \cite{mir,ali,wily} and in confined structures \cite{corna}
have shown that the Kondo peak near the Fermi energy splits in two for certain
parameters. Dias da Silva \textit{et al.} studied the spectral density at the QD 1 
$\rho_{d\sigma }(\omega )$ and magnetic susceptibility of the symmetric model
using numerical renormalization group (NRG). 
In agreement with previous studies, they find that when $\lambda $
increases beyond a certain critical value $\lambda _{c}$, the Kondo
resonance splits. In a similar way, for large enough 
$\lambda$ split peaks are observed if $U$ is smaller than a critical value 
$U_{c}$. In fact previous calculations for the spectral density of an
impurity inside a circular corral show split peaks in the non-interacting
case $U=0$ that turn into one peak as $U$ is increased.\cite{hvar} The
splitting condition is proposed to correspond approximately to the
equality $\sqrt{2}T_{K}=\Delta _{2}$, where $T_{K}$ is the Kondo
temperature. However, the results presented to support this reasonable
statement are limited. The spectral density is shown only for three
particular sets of parameters. Moreover, the ability of the NRG to describe
peaks far from the Fermi energy  might be questionable
because due to the logarithmic discretization, the resolution at a given
energy $\omega $ is proportional to $\omega -\epsilon _{F}$. In addition, the
spectral broadening affects the resolution.\cite{bulla}

In this paper we use perturbation theory in the Coulomb repulsion $U$ (PTU)
up to second order \cite{yos,hor} and a slave-boson mean-field approximation
(SBMFA) \cite{kr} to study in more detail the spectral density, the
conditions for the appearance of split peaks near $\epsilon _{F}$, and the
evolution of the energy scale $T_{K}$ with the parameters of the model. We
also compare with exact results in the limit $\Delta _{2}\rightarrow 0$, and
derived from the Friedel sum rule.\cite{lan} As explained in the next
section, both approximations take a simpler form in the symmetric Anderson
model considered, and the PTU works better in this case, allowing us to
describe new physics, and obtain reasonably robust results with
comparatively simple mathematics.

For the SBMFA we used a straightforward extension of the method proposed by
Kotliar and Ruckenstein to the Hubbard model.\cite{kr}

The paper is organized as follows. The model, the approximations (PTU and SBMFA)
and some exact results derived from Fermi liquid properties and the case
$\Delta_2=0$ are presented in Section II. Several figures illustrating the main
results are presented in Section III. Section IV is a short summary and
discussion.

\section{Hamiltonian, approximations and exact results}

\label{model}

\subsection{Model and equations for the spectral density}

The system is described by the following Hamiltonian:

\begin{equation}
H=H_{1}+H_{2}+H_{l}+H_{\lambda }+H_{V},  \label{h}
\end{equation}
where $H_{1}$ ($H_{2}$) describes the interacting (non-interacting) QD, $%
H_{l}$ the leads, and the last two terms are the hybridization of the
non-interacting QD with the interacting one and the leads respectively. In
standard notation

\begin{eqnarray}
H_{1} &=&E_{d}\sum_{\sigma }d_{\sigma }^{\dagger }d_{\sigma }+U
d_{\uparrow }^{\dagger }d_{\uparrow }d_{\downarrow }^{\dagger }d_{\downarrow
},  \nonumber \\
H_{2} &=&\epsilon _{a}\sum_{\sigma }a_{\sigma }^{\dagger }a_{\sigma },\text{ 
}H_{l}=\sum_{l\sigma }\epsilon _{l}c_{l\sigma }^{\dagger }c_{l\sigma }, 
\nonumber \\
H_{\lambda } &=&\lambda \sum_{\sigma }(d_{\sigma }^{\dagger }a_{\sigma }+%
\text{H.c.}),  \nonumber \\
H_{V} &=&\sum_{l\sigma }W_{l}(a_{\sigma }^{\dagger }c_{l\sigma }+\text{H.c.}%
).  \label{h2}
\end{eqnarray}
The sum $H_{2}+H_{l}+H_{V}$ is a non-interacting Hamiltonian which can be
put in the diagonal form $\sum_{k\sigma }\epsilon _{k}c_{k\sigma }^{\dagger
}c_{k\sigma }$ by means of a canonical transformation. In this basis the
Hamiltonian takes the form of the impurity Anderson model for a general band
structure and hybridization

\begin{equation}
H=H_{1}+\sum_{k\sigma }\epsilon _{k}c_{k\sigma }^{\dagger }c_{k\sigma
}+\sum_{k\sigma }V_{k}(d_{\sigma }^{\dagger }c_{k\sigma }+\text{H.c.}),
\label{h3}
\end{equation}
where $V_{k}=\lambda \{a_{\sigma },c_{k\sigma }^{\dagger }\}$.

The spectral density of electrons at the interacting QD is

\begin{equation}
\rho _{d\sigma }(\omega )=\frac{1}{2\pi }[G_{dd\sigma }(\omega -i\eta
)-G_{dd\sigma }(\omega +i\eta )],  \label{ro}
\end{equation}
where $\eta $ is a positive infinitesimal and calling $z=\omega +i\eta $ ($%
z=\omega -i\eta $), the retarded (advanced) Green's function at the 
interacting QD can be written in the form \cite{mir,lan}

\begin{equation}
G_{dd\sigma }(z)=\frac{1}{z-E_{d}-\Sigma _{sd\sigma }(z)-\Sigma _{dd\sigma
}(z)},  \label{gdd}
\end{equation}
where $\Sigma _{dd\sigma }(z)$ is the self-energy due to the
interaction $U$ and

\begin{equation}
\Sigma _{sd\sigma }(z)=\sum_{k}\frac{|V_{k}|^{2}}{z-\epsilon _{k}}.
\label{sd1}
\end{equation}
While this result holds for a general Hamiltonian of the form (\ref{h3}),
for the particular case of Eqs. (\ref{h}) and (\ref{h2}), using equations
of motion,\cite{mir} it can be shown that \cite{dias}

\begin{equation}
\Sigma _{sd\sigma }(z)=\lambda ^{2}G_{aa\sigma }^{0}(z),  \label{sd2}
\end{equation}
where $G_{aa\sigma }^{0}(z)$ is the Green's function of the non-interacting
QD in the absence of the interacting one (for a Hamiltonian $%
H_{2}+H_{l}+H_{V}$). The model assumes constant values for the matrix element 
$W_{l}$ and the density of
states of the leads $\rho $.\cite{dias} These assumptions are usually very
good approximations for the range of energies of interest in QD's. Calling $%
\Delta _{2}=\pi \rho |W|^{2}$, they lead to

\begin{equation}
G_{aa\sigma }^{0}(\omega \pm i\eta )=\frac{1}{\omega -\epsilon _{a}\pm
i\Delta _{2}}.  \label{gaa}
\end{equation}
In the following, as in Ref. \onlinecite{dias}, we take the origin of one-particle
energies at the Fermi energy ($\epsilon _{F}=0$), and take $\epsilon _{a}=0$%
, $\epsilon _{d}=-U/2$, corresponding to the symmetric Anderson model. In
summary, the model is equivalent to an impurity Anderson model with the
hybridization function

\begin{equation}
\Delta (\omega )=-\mathrm{Im}\Sigma _{sd\sigma }(\omega +i\eta )=\frac{\lambda
^{2}\Delta _{2}}{\omega ^{2}+\Delta _{2}^{2}}.  \label{dw}
\end{equation}
Note that in general, the real part of $\Sigma _{sd\sigma }$ can be obtained
from the relation

\begin{equation}
\Lambda (\omega )=\mathrm{Re}\Sigma _{sd\sigma }(\omega )=P\int \frac{\Delta
(\epsilon )d\epsilon }{\pi (\omega -\epsilon )}  \label{lw}
\end{equation}

If some approximation for $\Sigma _{dd\sigma }(z)$ is used, the above
equations (\ref{ro}), (\ref{gdd}), (\ref{sd2}) and (\ref{gaa}) 
define the spectral density.

\subsection{Approximations}

\subsubsection{Perturbations in $U$}

The starting point of the perturbation theory in $U$ (PTU) is a
non-interacting problem which includes some one-body potential so that the
effective on-site energy of the interacting electrons $E_{d}$ is modified to 
$E_{d\sigma }^{eff}$. A possible choice is the Hartree-Fock value $%
E_{d\sigma }^{eff}=E_{d}+U\langle d_{\bar{\sigma}}^{\dagger }d_{\bar{\sigma}}\rangle $. 
This one-body potential is compensated in the perturbation,
which includes it (with the opposite sign) in addition to the interaction term 
$U d_{\uparrow}^{\dagger }d_{\uparrow }d_{\downarrow }^{\dagger }d_{\downarrow }$. In
general, there are better choices for $E_{d\sigma }^{eff}$ which have been
used recently in several calculations of transport properties in nanoscopic
systems, including non-equilibrium situations and applied magnetic fields.\cite{none} 
However, for the symmetric Anderson model at equilibrium without
applied magnetic field 
(for which $\langle d_{\sigma }^{\dagger }d_{\sigma}\rangle =1/2$) these choices 
coincide with the Hartree-Fock result and $E_{d\sigma }^{eff}=E_{d}+U/2=\epsilon _{F}$. 
In this case, for a flat band
the theory is quantitatively correct up to $U\sim 8\Delta $, where $\Delta $
is the resonant level width.\cite{silver,dots} Working up to order $U^{2}$
we can write \cite{yos,hor,mir}

\begin{equation}
\lbrack G_{dd\sigma }(z)]^{-1}=[g_{dd\sigma }(z)]^{-1}-\Sigma _{dd\sigma
}^{(2)}(z),  \label{gd}
\end{equation}
where $g_{dd\sigma }(z)$ is the Green's function for the interaction treated
in Hartree-Fock (including contributions of first order in $U$ 
to $\Sigma _{dd\sigma }$)

\begin{equation}
g_{dd\sigma }(z)=\frac{1}{z-\Sigma _{sd\sigma }(z)},  \label{g0}
\end{equation}
and $\Sigma _{dd\sigma }^{(2)}(z)$ is the contribution to the self-energy of
second order in $U$ evaluated from a Feynman diagram involving the
analytical extension of the time ordered $g_{dd\sigma }$ to Matsubara
frequencies 
\begin{eqnarray}
\Sigma _{dd\sigma }^{(2)}(i\omega _{l},T) &=&U^{2}T\sum_{m}g_{dd\sigma
}(i\omega _{l}-i\nu _{m})\chi _{\bar{\sigma}}(i\nu _{m}),  \label{sigma} \\
\chi _{\sigma }(i\nu _{m}) &=&-T\sum_{n}g_{dd\sigma }(i\omega
_{n})g_{dd\sigma }(i\omega _{n}+i\nu _{m}),  \label{xi}
\end{eqnarray}
where $\omega _{n}=(2n+1)\pi T$ and $\nu _{m}=2m\pi T$.

Our task is simplified because $\chi _{\sigma }$ can be calculated
analytically due to the fact that $g_{dd\sigma }$ can be expressed as a sum
of two simple fractions 
\begin{equation}
g_{dd\sigma }(\omega +i\eta )=\frac{\omega +i\Delta _{2}}{\omega
^{2}+i\Delta _{2}\omega -\lambda ^{2}}=\sum_{l=1}^{2}\frac{a_{l}}{\omega
-r_{l}+i\delta _{l}}.  \label{gdp}
\end{equation}
This is the retarded Green's function. In order to evaluate perturbative
diagrams, like Eqs. (\ref{sigma}), (\ref{xi}), one needs the analytical
extension of the time ordered Green function to imaginary frequencies $%
\omega \rightarrow i\omega _{n}$ \cite{mahan,mir}: 
\begin{equation}
g_{dd\sigma }(i\omega _{n})=\sum_{l}\frac{a_{l}+\overline{a_{l}}+(a_{l}-%
\overline{a_{l}})\text{sgn}(\omega _{n})}{2\left[ i\omega _{n}+i\delta _{l}%
\text{sgn}(\omega _{n})-r_{l}\right] },  \label{gdn}
\end{equation}
where $\overline{a_{l}}$ is the complex conjugate of $a_{l}$ and sgn$(x)$ is
the sign of $x$.

The sums over Matsubara frequencies, Eqs. (\ref{sigma}), (\ref{xi}) have
been done in Ref. \onlinecite{mir}. The latter can be expressed in
terms of the digamma function $\Psi (z)$.\cite{abra} The final result for
the retarded quantities is 
\begin{eqnarray}
\Sigma _{dd\sigma }^{(2)}(\omega ) &=&\frac{U^{2}}{\pi }\int_{-\infty
}^{\infty }dy\ [\frac{1}{e^{y/T}-1}g_{dd\sigma }(\omega -y)\mathrm{Im}\chi _{%
\bar{\sigma}}(y)+  \nonumber \\
&&+\frac{1}{e^{y/T}+1}\mathrm{Im}g_{dd\sigma }(-y)\ \chi _{\bar{\sigma}%
}(\omega +y)],  \label{sigf}
\end{eqnarray}

where 
\begin{eqnarray}
\chi _{\sigma }(\omega ) &=&-\frac{i}{2\pi }\ \sum_{lm}\left( A_{lm}(\omega
)-B_{lm}(\omega )\right) \Psi _{l}(0)+  \nonumber \\
&&+\left( \overline{B_{lm}}(-\omega )-\overline{A_{lm}}(-\omega )\right) 
\overline{\Psi _{l}}(0)  \nonumber \\
&&-\left( B_{lm}(\omega )+A_{lm}(-\omega )\right) \Psi _{l}(\omega ) 
\nonumber \\
&&-\left( \overline{B_{lm}}(-\omega )+\overline{A_{lm}}(\omega )\right) 
\overline{\Psi _{l}}(-\omega ),  \label{xif}
\end{eqnarray}

with

\begin{eqnarray}
A_{lm}(\omega ) &=&\frac{a_{l}\ a_{m}}{\omega +r_{l}-r_{m}-i\delta
_{l}+i\delta _{m}},  \nonumber \\
B_{lm}(\omega ) &=&\frac{a_{l}\ \overline{a_{m}}}{\omega
-r_{l}+r_{m}+i\delta _{l}+i\delta _{m}},  \nonumber \\
\Psi _{l}(\omega ) &=&\Psi \left( \frac{1}{2}+\frac{\delta
_{l}+i(r_{l}-\omega )}{2\pi T}\right) .  \label{aux}
\end{eqnarray}
Eqs. (\ref{ro}), (\ref{sd2}), (\ref{gaa}), (\ref{gd}), (\ref{g0}) 
and (\ref{sigf}) to (\ref{aux}) define the
spectral density of the interacting dot within PTU.

\subsubsection{Slave bosons}

The basic idea of the slave boson formalism of Kotliar and Ruckenstein \cite
{kr} is to enlarge the Fock space to include bosonic states which correspond
to each state in the fermionic description at the interacting QD. The vacuum
state at this site is represented as $e^{\dagger }\left\vert 0\right\rangle $%
, where $e^{\dagger }$ is a bosonic creation operator corresponding to the
empty QD; similarly $s_{\sigma }^{\dagger }f_{\sigma }^{\dagger }\left\vert
0\right\rangle $ represents the singly occupied state $d_{\sigma }^{\dagger
}\left\vert 0\right\rangle $, where $s_{\sigma }^{\dagger }$ is a bosonic
operator for singly occupied sites with spin $\sigma $, and $f_{\sigma
}^{\dagger }$ is a fermion operator. The doubly occupied site $d_{\uparrow
}^{\dagger }d_{\downarrow }^{\dagger }\left\vert 0\right\rangle $ is
represented as $b^{\dagger }f_{\uparrow }^{\dagger }f_{\downarrow }^{\dagger
}\left\vert 0\right\rangle $ In this way the interaction term can be
expressed in terms of boson operators as $Ub^{\dagger }b$ and the
interactions between fermions disappear from the Hamiltonian. In other
words, the fermion operator $d_{\sigma }^{\dagger }$ is expressed in terms
of fermion operators that do not interact between them as $d_{\sigma
}^{\dagger }=(b^{\dagger }s_{\bar{\sigma}}+s_{\sigma }^{\dagger }e)f_{\sigma
}^{\dagger }$. The bosonic operators should satisfy the following constraints

\begin{eqnarray}
e^{\dagger }e+\sum_{\sigma }s_{\sigma }^{\dagger }s_{\sigma }+b^{\dagger }b
&=&1  \nonumber \\
s_{\sigma }^{\dagger }s_{\sigma }+b^{\dagger }b &=&f_{\sigma }^{\dagger
}f_{\sigma }.  \label{cons}
\end{eqnarray}
Introducing Lagrange multipliers for these constraints, 
the Hamiltonian takes the form

\begin{eqnarray}
H &=&H_{2}+H_{l}+H_{V}+E_{d}\sum_{\sigma }f_{\sigma }^{\dagger }f_{\sigma
}+Ub^{\dagger }b  \nonumber \\
&&+\lambda \sum_{\sigma }\left[ R_{\sigma }(b^{\dagger }s_{\bar{\sigma}%
}+s_{\sigma }^{\dagger }e)L_{\sigma }f_{\sigma }^{\dagger }a_{\sigma }+\text{%
H.c.}\right]  \nonumber \\
&&+\mu _{\sigma }(s_{\sigma }^{\dagger }s_{\sigma }+b^{\dagger }b-f_{\sigma
}^{\dagger }f_{\sigma })  \nonumber \\
&&-\mu (e^{\dagger }e+\sum_{\sigma }s_{\sigma }^{\dagger }s_{\sigma
}+b^{\dagger }b-1).  \label{hb}
\end{eqnarray}
Here, the factors

\begin{eqnarray}
R_{\sigma } &=&(1-e^{\dagger }e-s_{\sigma }^{\dagger }s_{\sigma })^{-1/2}\ ,
\nonumber \\
L_{\sigma } &=&(1-s_{\sigma }^{\dagger }s_{\sigma }-b^{\dagger }b)^{-1/2},
\label{r}
\end{eqnarray}
are equivalent to one when treated exactly, but they are introduced in such
a way that in the slave-boson mean-field approximation (SBMFA) the correct
result in the non-interacting case is reproduced.

In the SBMFA, all the boson operators are replaced by numbers and the free
energy is minimized with respect to them. In the symmetric Anderson model,
without applied magnetic field the problem simplifies considerably. In this
case $s^{2}$ is independent of spin and $e^{2}=b^{2}$. Also $\langle
d_{\sigma }^{\dagger }d_{\sigma }\rangle =\langle f_{\sigma }^{\dagger
}f_{\sigma }\rangle =1/2$. From here $E_{d}-\mu _{\sigma }=\epsilon _{F}=0$
and using Eqs. (\ref{cons}) $s^{2}=1/2-b^{2}$ and only one independent
variable remains. Also $R_{\sigma }=L_{\sigma }=\sqrt{2}$. The change in
free energy due to the impurity can be written as \cite{hew,3do}

\begin{eqnarray}
\Delta F=-\frac{2}{\pi }\mathrm{Im}\int_{-\infty }^{+\infty }f(\omega )\ln
G_{ff\sigma }(\omega +i\eta )\ d\omega \ +Ub^{2} \nonumber \\
+2 \mu_{\sigma}(s^{2}+b^{2}),  \label{f}
\end{eqnarray}
where $f(\omega )$ is the Fermi function and $G_{ff\sigma }(\omega )$ is the
Green's function of the $f$ operators for any spin. The Hamiltonian takes
the same form as that of a non-interacting problem with $E_{d}=\epsilon
_{F}=0 $ and renormalized hopping $\tilde{\lambda}=R_{\sigma }(b^{\dagger
}s_{\bar{\sigma}}+s_{\sigma }^{\dagger }e)L_{\sigma }\lambda $, that in our
case simplifies to

\begin{equation}
\tilde{\lambda}=4b\sqrt{1/2-b^{2}}\lambda .  \label{lt}
\end{equation}
Then, as in the previous subsection

\begin{equation}
G_{ff\sigma }(\omega +i\eta )=\frac{\omega +i\Delta _{2}}{\omega
^{2}+i\Delta _{2}\omega -\tilde{\lambda}^{2}}.  \label{gff}
\end{equation}
Decomposing this expression in simple fractions [as in Eq. (\ref{gdp})] and
replacing in Eq. (\ref{f}), the integral can be evaluated analytically at
zero temperature. The result should be separated in two cases depending on
the sign of $2\tilde{\lambda}-\Delta _{2}$. 
Except for irrelevant constants, the result is

\begin{eqnarray}
\Delta F &=&\frac{\Delta _{2}}{\pi }\{\ln (1+r^{2})+r(2\arctan r^{-1}-\pi
)\}+Ub^{2}\text{ }  \nonumber \\
\text{for  }x &\geq &1\text{, and}  \nonumber \\
\Delta F &=&\frac{\Delta _{2}}{\pi }\{(1-r)\ln (1-r)+(1+r)\ln (1+r) \}+Ub^{2},  \nonumber \\
\text{for  }x &\leq &1\text{, with}  \nonumber \\
r &=&\sqrt{|x^{2}-1|},\text{ }x=2\tilde{\lambda}/\Delta _{2}.  \label{df2}
\end{eqnarray}

Minimizing $\Delta F$ defined by Eqs. (\ref{df2}) and (\ref{lt}), one
obtains a transcendental equation for $b$. After solving this, a
characteristic energy scale or Kondo temperature can be defined by the gain
in energy with respect to the unhybridized case:

\begin{equation}
T_{K}=\Delta F(\tilde{\lambda}_{\min })-\Delta F(0),  \label{tk}
\end{equation}
where $\tilde{\lambda}_{\min }$ is the value of $\tilde{\lambda}$ evaluated
with Eq. (\ref{lt}) for the value of $b$ that minimizes the energy.

The spectral density for real $d$ electrons near the Fermi energy becomes,
using $d_{\sigma }^{\dagger }=R_{\sigma }(b^{\dagger }s_{\bar{\sigma}%
}+s_{\sigma }^{\dagger }e)L_{\sigma }f_{\sigma }^{\dagger }$ in the SBMFA

\begin{equation}
G_{dd\sigma }(\omega +i\eta )\simeq (\tilde{\lambda}_{\min }/\lambda
)^{2}G_{ff\sigma }(\omega +i\eta ).  \label{gddb}
\end{equation}
In the resulting spectral density $\rho _{d\sigma }(\omega )$ [calculated
from Eq. (\ref{ro})], the ``charge transfer'' peaks near $E_{d}$ and $%
E_{d}+U $ are lost, as explained in more detail in the next section.

From the change of sign of $\partial ^{2}\rho _{d\sigma }(\omega )/\partial
\omega ^{2}$ evaluated at the Fermi energy, one finds that (in a similar way
as in the non-interacting case \cite{dias}) the critical condition for the
appearance or disappearance of split peaks near the Fermi energy is

\begin{equation}
\sqrt{2}\tilde{\lambda}_{\min }=\Delta _{2}\text{.}  \label{ltc}
\end{equation}
Combining this equation with Eq. (\ref{lt}) and the minimization condition $%
\partial \Delta F/\partial b=0$, we find in order to have split peaks, it is
necessary that $\lambda >\Delta _{2}/\sqrt{2}$ and that $U<U_{c}$, where

\begin{equation}
U_{c}=\frac{8\lambda ^{2}}{\Delta _{2}}\left( 1-\frac{\Delta _{2}^{2}}{%
2\lambda ^{2}}\right) ^{1/2}.  \label{uc}
\end{equation}
This analytical result can be inverted to give the minimum value of $\lambda 
$ required to have split peaks for fixed $U$ and $\Delta _{2}:$

\begin{equation}
\lambda _{c}=\frac{U}{4\left( \left( 1+\frac{U^{2}}{4\Delta _{2}^{2}}\right)
^{1/2}-1\right) ^{1/2}}.  \label{lc}
\end{equation}

\subsection{Exact results}

\subsubsection{Fermi liquid properties}

It is known that at the Fermi energy, the imaginary part of the self energy
due to interaction vanishes in a Fermi liquid.\cite{lutt} Using 
$\mathrm{Im}\Sigma _{dd\sigma }(\epsilon _{F})=0$, and Eqs. (\ref{ro}) and (\ref{gdd})
with $\Sigma _{sd\sigma }(\omega \pm i\eta )=\Lambda _{\sigma }(\omega )\mp
i\Delta _{\sigma }(\omega )$, the spectral density at the Fermi energy can
be written in the form

\begin{equation}
\rho _{d\sigma }(\epsilon _{F})=\frac{\cos ^{2}\varphi _{\sigma }}{\pi
\Delta _{\sigma }(\epsilon _{F})},  \label{l1}
\end{equation}
where

\begin{equation}
\varphi _{\sigma }=\arctan \left( \frac{E_{d}+\Sigma _{dd\sigma }(\epsilon
_{F})+\Lambda _{\sigma }(\epsilon _{F})}{\Delta _{\sigma }(\epsilon _{F})}%
\right) .  \label{fi}
\end{equation}
In addition, for the general Anderson model, the Friedel sum rule is valid \cite{lan}

\begin{eqnarray}
\varphi _{\sigma }=\pi \left( \langle d_{\sigma }^{\dagger }d_{\sigma }\rangle -%
\frac{1}{2} \right) \nonumber \\
+\mathrm{Im}\int_{-\infty }^{\epsilon _{F}}d\omega G_{dd\sigma
}(\omega +i\eta )\frac{\partial \Sigma _{sd\sigma }(\omega +i\eta )}{%
\partial \omega }.  \label{fsr}
\end{eqnarray}

As explained below, for the symmetric Anderson model without applied magnetic field, with the
choice $\epsilon_{F}=0$, 
it can be shown that the real (imaginary) part of $G_{dd\sigma
}(\omega +i\eta )$ is odd (even).
Then Eq. (\ref{gdd}) implies 
$\mathrm{Re}[E_{d}+\Sigma _{sd\sigma }(0)+\Sigma _{dd\sigma }(0)]=0$.
From Eq. (\ref{fi}) $\varphi _{\sigma }=0$ and from Eq. (\ref{l1})

\begin{equation}
\rho _{d\sigma }(0)=\frac{1}{\pi \Delta (0)}.  \label{ro0}
\end{equation}

The symmetry properties of $G_{dd\sigma }(\omega +i\eta )$ can be
demonstrated using the Lehman representation of the Green's function \cite
{mahan} and the fact that the symmetric Anderson model is invariant under
the transformation $T:d_{\sigma }^{\dagger }\rightarrow d_{\sigma }$, $c_{%
\mathbf{k}\sigma }^{\dagger }\rightarrow -c_{\mathbf{k}^{\prime }\sigma }$,
with $\epsilon _{\mathbf{k}^{\prime }}=-\epsilon _{\mathbf{k}}$. Calling $%
\Omega $ the thermodynamic potential and $|n\rangle $ a complete set of
eigenstates, we can write \cite{mahan}

\begin{equation}
G_{dd\sigma }(\omega +i\eta )=e^{\beta \Omega }\sum_{n,m}|\langle
n|d_{\sigma }^{\dagger }|m\rangle |^{2}\frac{e^{-\beta E_{n}}+e^{-\beta
E_{m}}}{\omega +E_{n}-E_{m}+i\eta }.  \label{leh}
\end{equation}
Because of symmetry one has $\langle n|d_{\sigma }^{\dagger }|m\rangle
=\langle Tn|Td_{\sigma }^{\dagger }T^{\dagger }|Tm\rangle $ $=\langle
Tn|d_{\sigma }|Tm\rangle $ 
$=\overline{\langle Tm|d_{\sigma }^{\dagger
}|Tn\rangle}$, and the eigenstates $|m\rangle $ and $|Tm\rangle $ have the
same energy. Then changing the labels of the sum above $|n\rangle $ by 
$|Tm\rangle $ and $|m\rangle $ by $|Tn\rangle $

\[
G_{dd\sigma }(-\omega -i\eta ) \\
=e^{\beta \Omega }\sum_{n,m}|\langle
Tm|d_{\sigma }^{\dagger }|Tn\rangle |^{2}\frac{e^{-\beta E_{n}}+e^{-\beta
E_{m}}}{-\omega +E_{m}-E_{n}-i\eta }, 
\]
and using the symmetry property of the matrix elements indicated above, one
finds

\begin{equation}
G_{dd\sigma }(-\omega -i\eta )=-G_{dd\sigma }(\omega +i\eta ).  \label{sym}
\end{equation}

We have verified that Eqs. (\ref{fsr}) and (\ref{ro0}) are satisfied by the
approximations (PTU and SBMFA) presented above. For the PTU the second
member of Eq. (\ref{fsr}) has been evaluated numerically and we find that it
is zero within the accuracy of the computer. Instead, within NRG, Eq. (\ref{ro0}) 
is satisfied only approximately.\cite{bulla}

\subsubsection{The limit $\Delta _{2}\rightarrow 0$}

This limit coincides with the so called atomic limit of the Anderson model,
reported previously in Appendix C of Ref. \onlinecite{hew} and in Ref. \onlinecite{allub}%
. We describe here the main results for the symmetric Anderson model.

The ground state is a two-particle singlet with energy $E_{g}=E_{t}-T_{K}$,
where $E_{t}=-U/2$ is the energy of the triplet state and the characteristic
energy $T_{K}$ is

\begin{equation}
T_{K}=\frac{U}{4}\left[ \sqrt{1+\left( \frac{8\lambda }{U}\right) ^{2}}%
-1\right] .  \label{tkd0}
\end{equation}
This energy coincides with the gain in energy due to hybridization, Eq. (\ref
{tk}). Note that for $U\rightarrow \infty $, $T_{K}=8\lambda ^{2}/U$ in
contrast to the result for a flat band $T_{K}\simeq D\exp [-\pi U/4\Delta ].$

The spectral density is easily calculated using Eq. (\ref{ro}) and the
Lehman representation Eq. (\ref{leh}) of the Green's function. For later
use, we display here the result at $T=0$

\begin{eqnarray}
\rho _{d\sigma }(\omega ) &=&A[\delta (\omega -E_{A})+\delta (\omega +E_{A})]
\nonumber \\
&&+(1/2-A)[\delta (\omega -E_{B})+\delta (\omega +E_{B})],  \label{rol0}
\end{eqnarray}
where $\pm E_{A}$ are the positions of the peaks nearest to the Fermi level
and $A$ their weight, while $\pm E_{B}$ and $1/2-A$ are the position and
weights of the \textquotedblleft charge transfer\textquotedblright\ peaks
nearer to $\pm U/2$. The values of these quantities are

\begin{eqnarray}
E_{A} &=&\frac{U}{4}\left[ \sqrt{1+\left( \frac{8\lambda }{U}\right) ^{2}}
-\sqrt{1+\left( \frac{4\lambda }{U}\right) ^{2}}\right],  \nonumber \\
E_{B} &=&\frac{U}{4}\left[ \sqrt{1+\left( \frac{8\lambda }{U}\right) ^{2}}
+\sqrt{1+\left( \frac{4\lambda }{U}\right) ^{2}}\right],  \nonumber \\
A &=&\frac{1}{8}\left[ \sqrt{(1+y_{4})(1-y_{8})}+\sqrt{(1-y_{4})(1+y_{8})}
\right] ^{2},  \nonumber \\
y_{j} &=&\left[ 1+(j\lambda /U)^{2}\right] ^{-1/2}.  \label{w}
\end{eqnarray}
For $U\rightarrow \infty $, they simplify to

\begin{eqnarray}
E_{A} &=&6\lambda ^{2}/U  \nonumber \\
E_{B} &=&U/2+10\lambda ^{2}/U  \nonumber \\
A &=&18\lambda ^{2}/U^{2}.  \label{winf}
\end{eqnarray}
When the temperature is increased, more peaks appear in the spectral density
and the weight $A$ of the peaks nearer to the Fermi energy decreases.

\section{Approximate results}

\subsection{General aspect of the spectral density}

\begin{figure}[tbh]
\includegraphics[width=.9\linewidth]{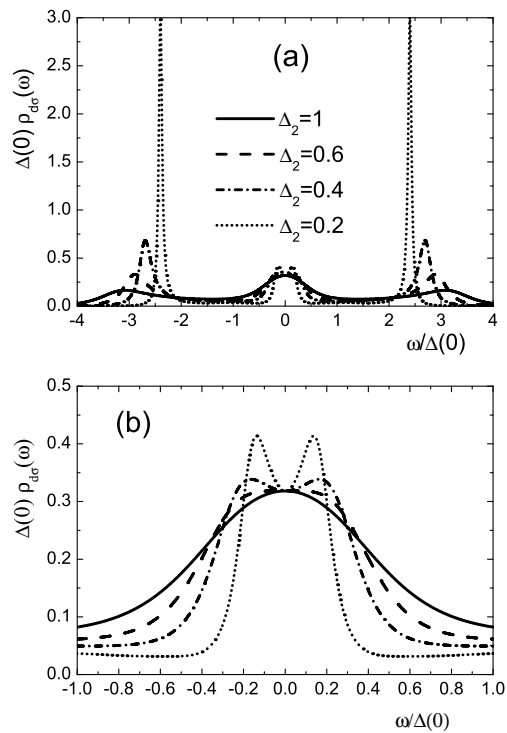} 
\caption{Spectral density of states for $U=4$ and several 
values of $\Delta_2/\Delta(0)$ keeping $\lambda= \sqrt{\Delta_2 \Delta(0)}$.}
\label{per1}
\end{figure}

In Fig. \ref{per1} we show the spectral density calculated with PTU. We have
fixed $\Delta (0)=\lambda ^{2}/\Delta _{2}$ as the unit of energy. Then
according to Eq. (\ref{ro0}) the spectral density at the Fermi level is $%
1/\pi $, for all sets of parameters. On can see that the PTU satisfies this
condition derived from Fermi liquid properties. In addition to this
observation, there are two other noticeable properties that can be observed
in the figure, in comparison with the case of constant $\Delta (\omega ).$
One is the presence of split peaks near the Fermi level for large enough 
$\lambda $ and $U$, already observed before \cite{hvar,dias} as discussed in
the introduction. The other is that the \textquotedblleft charge
transfer\textquotedblright\ peaks at $E_{d}$ and $E_{d}+U$ are unusually
high and narrow, particularly when split peaks appear. We begin discussing
the latter fact. This has not been noticed by previous NRG calculations.\cite
{dias} We believe that this is due to the lack of resolution of the NRG
for structures that are far form the Fermi energy. We remind the reader that
for constant $\Delta (\omega )$ these side peaks are broad and of smaller
amplitude than the only central peak (the so called Kondo peak). In
particular, in the symmetric case $E_{d}=-U/2$, while the width of the Kondo
peak (the half width at half maximum) is of the order of the Kondo scale $%
T_{K}$ and its height is $1/(\pi \Delta )$, the width of the side peaks is
of order $2\Delta $ and their height is near $1/(4\pi \Delta )$. A
discussion of the effects that lead to the extra broadening of these peaks
was presented by Logan et al.\cite{logan} These authors also show that
qualitatively these peaks can be understood using an alloy analog approach
(AAA) that consists in replacing the system by a non interacting system in
which the on-site energy of the localized electrons is either $E_{d}$ or $%
E_{d}+U$ with equal probability. This approach misses the Kondo peak but
describes surprisingly well the charge transfer peaks for large $U$.\cite
{logan} In our case, the AAA also shows high and narrow peaks for small
enough $\Delta _{2}$ and the reason is that $\Delta (\omega )$ is
considerably smaller for $\omega =E_{d}$ or $\omega =E_{d}+U$ than at the
Fermi energy $\Delta (0)$. While this analysis gives confidence to our
results, the reader might still wonder if the aspect of the side peaks is
due to a shortcoming of the PTU. For constant $\Delta (\omega )$ (large $%
\Delta _{2}$) the approach is quantitatively valid for $U\lesssim 8\Delta .$%
\cite{silver,dots} \ In the following subsection we analyze the precision of
the method for small $\Delta _{2}$.

Near the Fermi energy, we see that as the ratio $\lambda /\Delta _{2}$
increases, first the Kondo peak narrows, then its splits in two and with
further increase in $\lambda /\Delta _{2}$ the split peaks became narrower
and higher. They also tend to move towards the Fermi energy but they never
merge into one again (for $\lambda \gg \Delta_2$, the exact results of Section II.2 become qualitatively valid).

\begin{figure}[tbh]
\includegraphics[width=.9\linewidth]{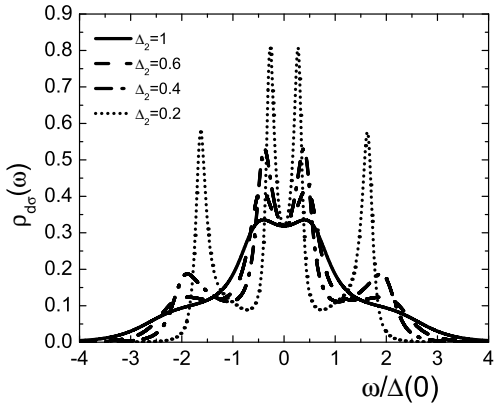} 
\caption{Spectral density of states for $U=2$ and several 
values of $\Delta_2/\Delta(0)$ keeping $\lambda= \sqrt{\Delta_2 \Delta(0)}$.}
\label{per2}
\end{figure}

As expected, if $U$ is lowered, the values of $\lambda /\Delta _{2}$ needed
to have split peaks are smaller. This can be seen comparing Fig. \ref{per1} 
($U=4\Delta $) with Fig. \ref{per2} ($U=2\Delta $), or Fig. 2 of a previous
work on the quantum mirage effect,\cite{hvar} in which split peaks are
obtained only in the non-interacting case of the model used.

\subsection{Variation with $\Delta _{2}$}

\begin{figure}[tbh]
\includegraphics[width=.9\linewidth]{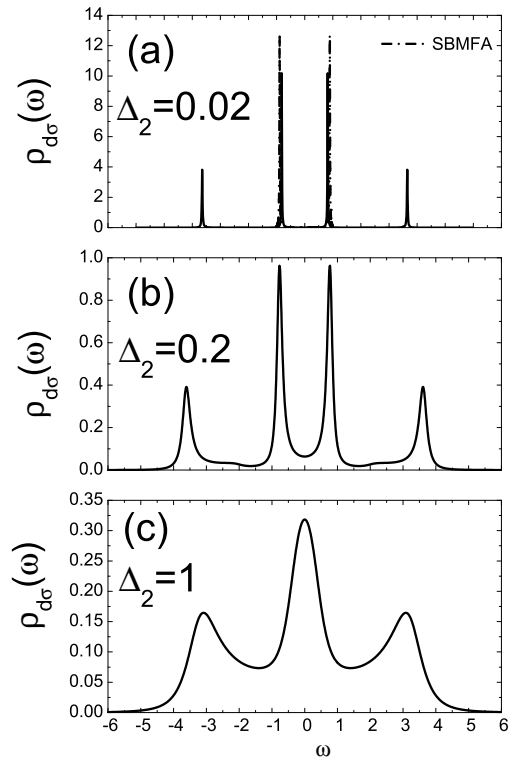} 
\caption{Spectral density of states for $U=4$, $\lambda=1$ and several 
values of $\Delta_2$.}
\label{del1}
\end{figure}

\begin{figure}[tbh]
\includegraphics[width=.9\linewidth]{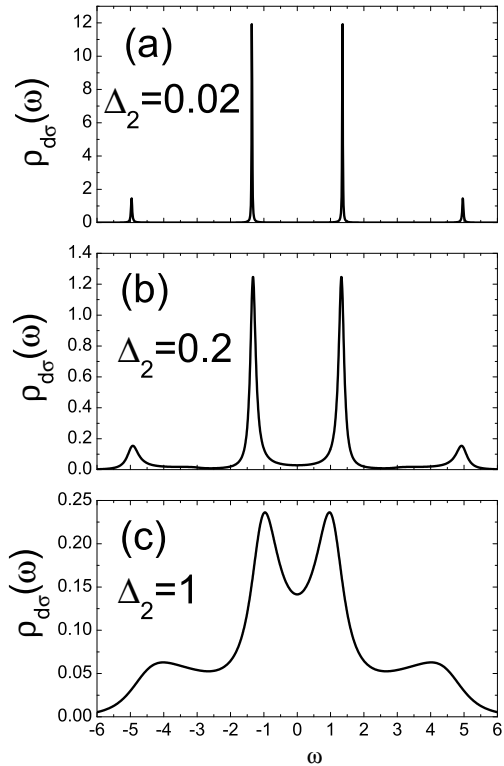} 
\caption{Spectral density of states for $U=4$, $\lambda=1.5$ and several 
values of $\Delta_2$.
}
\label{del2}
\end{figure}

In Figs. \ref{del1} and \ref{del2} we show the evolution of the spectral
density calculated with PTU as $\Delta _{2}$ is increased, starting from
very small values. As $\Delta_2$ increases, all peaks broaden and those
nearer to the Fermi energy approach each other, merging into one for large
enough $\Delta _{2}$. This evolution is similar to that reported previously
in Fig. 2 of Ref. \onlinecite{ali} as a function of the width of the 
resonances in the Anderson model for an impurity inside a quantum corral.
In Fig. \ref{del1} we choose $\lambda $ as the unit of
energy and the evolution ends with only one peak at the Fermi energy when 
$\Delta _{2}$ reaches 1. Increasing $\lambda $ to 1.5, the magnitude of the
splitting increases and the peaks near the Fermi energy are higher and
remain split for $\Delta _{2}=1$. 

For $\Delta _2=0.02$ we have integrated numerically the spectral density $%
\rho _{d\sigma }(\omega )$ below each peak to compare with the exact results
given by Eqs. (\ref{w}). From the energy of the maxima of $\rho _{d\sigma
}(\omega )$ ($\pm E_{A}$, $\pm E_{B}$) and the resulting values of the peak
weights ($A$ and $1/2-A$) we obtain $E_{A}=0.82$, 
$E_{B}=3.65 $, and $A=0.325$ for the
parameters of Fig. \ref{del1} and $E_{A}=1.36$, $E_{B}=4.96$, and $A=0.401$ for the
parameters of Fig. \ref{del2}. The corresponding exact values
for $\Delta _{2}=0$ given by Eqs. (\ref{w}) differ in less than 0.004 from
those given above. This shows that the PTU works very well for small $\Delta
_{2}$ and provides confidence to the results presented above.

In Fig. \ref{del1} (a), the result of the SBMFA is also shown for
comparison. This approximation also gives qualitatively correct results for
the two peaks nearer to the Fermi energy for small $\Delta_2$, but 
quantitatively the PTU is superior in this limit.

\begin{figure}[tbh]
\includegraphics[width=.9\linewidth]{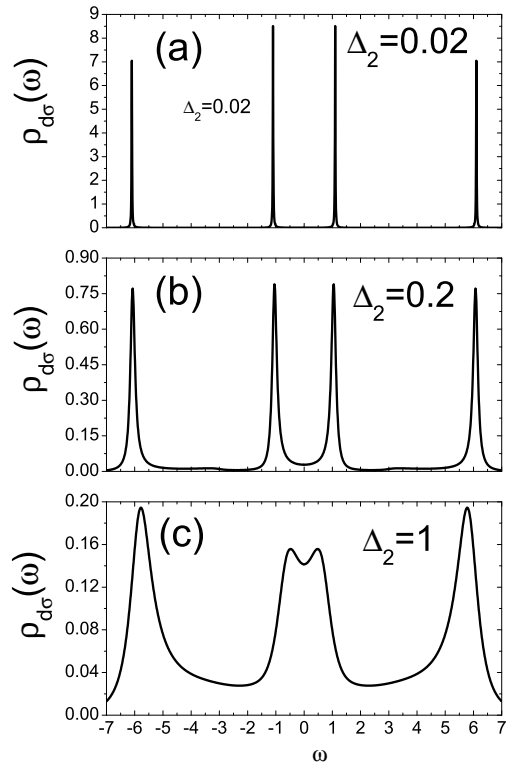} 
\caption{Spectral density of states for $U=8$, $\lambda=1.5$ and several 
values of $\Delta_2$.
}
\label{del3}
\end{figure}

In Fig. \ref{del3} we show the effect of a larger $U$. 
When comparing with Fig. 
\ref{del2}, we see that the peaks near the Fermi energy become closer again
and spectral weight is transferred to the side peaks. In this case, we
obtain within PTU for $\Delta _{2}=0.02$, 
$E_{A}=1.09$, $E_{B}=6.10$, and $A=0.261$ .
These values differ from the exact results for $\Delta _{2}=0$ in less
than 1\%.

\subsection{Description of the splitting in terms of quasiparticles.}

While the limit $\Delta _{2}\rightarrow 0$ provides a qualitative
description of split peaks near the Fermi energy, a more precise description
can be given in terms of weakly interacting quasiparticles, for which the
analytical results of a non-interacting systems with renormalized parameters
gives an accurate description as a starting point, as we show.

\begin{figure}[tbh]
\includegraphics[width=.9\linewidth]{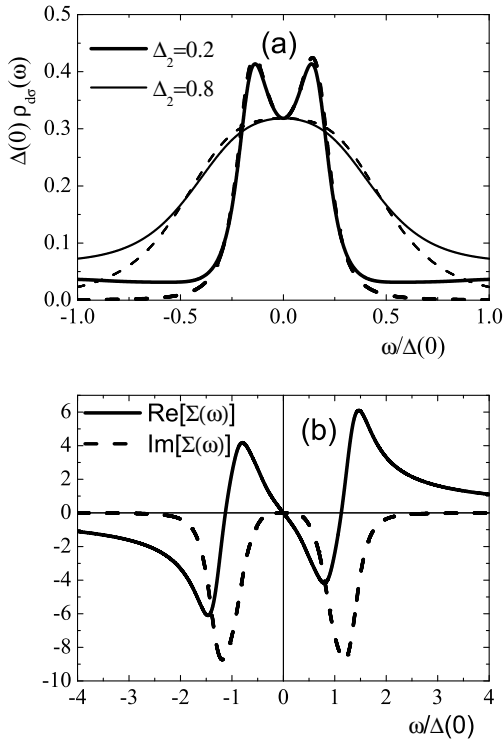} 
\caption{(a) Spectral density of states near the Fermi level 
for two values of $\Delta_2$ keeping $\Delta(0)=1$. Full lines: $U=4$, 
dashed lines: renormalized non-interacting case.
(b) Self energy for $\Delta_2=0.2$}
\label{zf}
\end{figure}

Hewson has noticed that near the Fermi energy, it is possible to reformulate
the PTU in terms of quasiparticles, for which the residual interactions are
smaller.\cite{rpt} The starting point of renormalized perturbation theory,
as in Fermi liquid theory, is an expansion of the self energy $\Sigma
_{dd\sigma }(\omega +i\eta )$ around the Fermi energy (zero in our case).
Since $\mathrm{Im}\Sigma _{dd\sigma }(\omega +i\eta )\sim \omega ^{2}$ near
the Fermi energy [see Fig. \ref{zf} (b)], to linear order in $\omega $ we can
approximate Eq. (\ref{gdd}) as

\begin{eqnarray}
G_{dd\sigma }(\omega ) &\simeq &\frac{1}{\omega (1-\partial \mathrm{Re}\Sigma
_{dd\sigma }/\partial \omega )-E_{d}^{\prime }-\Sigma _{sd\sigma }(\omega )}
\nonumber \\
&=&\frac{z}{\omega -zE_{d}^{\prime }-z\Sigma _{sd\sigma }(\omega )},
\label{gz}
\end{eqnarray}
where $E_{d}^{\prime }=E_{d}+\Sigma _{dd\sigma }(0)$ (zero in our case) and

\begin{equation}
z=\frac{1}{(1-\partial \mathrm{Re}\Sigma _{dd\sigma }/\partial \omega )}.
\label{z}
\end{equation}
The interactions however still modify this picture and a renormalized self
energy should be introduced.\cite{rpt} However, we find that Eq. (\ref{gz})
with $z$ calculated numerically from the slope of $\mathrm{Re}\Sigma
_{dd\sigma }$ leads already a very good description of the spectral density
near the Fermi energy. This is shown in Fig. \ref{zf} for two different 
sets of parameters, one leading to split peaks and the other not. The
discrepancies between the complete PTU results and this approximation are
appreciable only sufficiently far from the Fermi energy, where the influence
of the side peaks, ignored in Eqs. (\ref{gz}) cannot be neglected. The resulting 
values of $z$ for $\Delta_2/\Delta(0)$ from 0.2 to 1 in steps of 0.2 with
$\lambda= \sqrt{\Delta_2 \Delta(0)}$
are 0.20, 0.28, 0.33, 0.36 and 0.38 respectively.   

Note that this approximation is equivalent to use a non-interacting model
with renormalized hybridization $\tilde{\lambda}=\sqrt{z}\lambda $, [see
Eqs. (\ref{dw}) and (\ref{lw})] and decrease the resulting $G_{dd\sigma
}(\omega +i\eta )$ by a factor $z<1$. This is very similar to the SBMFA [see
Eqs. (\ref{lt}) and (\ref{gddb})], except for the fact that in the latter,
the renormalized value of $\tilde{\lambda}$ is obtained by a minimization of
the energy and not by deriving an approximation to the self energy. Thus,
the results presented above can be regarded as a support to the qualitative
validity of the SBMFA to describe the spectral density $\rho _{d\sigma
}(\omega )$ near the Fermi energy. A comparison between results of PTU and
SBMFA for $\rho _{d\sigma }(\omega )$ is presented in subsection F.

\subsection{Critical conditions for split peaks}

\begin{figure}[tbh]
\includegraphics[width=.9\linewidth]{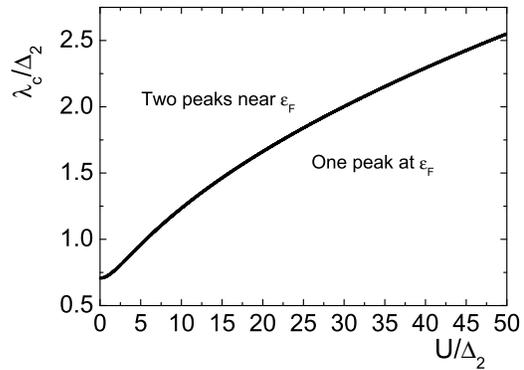} 
\caption{Boundary separating the region of parameters for which
split peaks in the spectral density are present.}
\label{lcf}
\end{figure}

Within the SBMFA we have obtained analytical results [see Eqs. (\ref{uc})
and (\ref{lc})] for relations between the parameters when split peaks just
appear. This leads to a kind of ``phase diagram'' for the appearance of
split peaks. The boundary line is shown in Fig. \ref{lcf}. Split peaks are
expected for larger values of $\lambda $ or smaller values of $U$, the
upper left of the curve. This results should be regarded as qualitative.
Comparison with Figs. \ref{per1} and \ref{per2} show that slightly larger $%
\lambda _{c}$ are expected within PTU. In fact, while using the factors $%
R_{\sigma }$ and $L_{\sigma }$[see Eqs. (\ref{r})] in the SBMFA leads to the
correct results in the non-interacting limit $U\rightarrow 0$, these roots
should be eliminated to obtain the right exponential dependence of $T_{K}$
in the limit $U\rightarrow \infty $ for constant $\Delta (\omega )$.\cite
{3do} This modification increases $\lambda _{c}$ by a factor 2.
Therefore, we expect that the actual value of $\lambda _{c}$ is larger than
that indicated in Fig. \ref{lcf}, particularly for larger values of $U$.

\begin{figure}[tbh]
\includegraphics[width=.9\linewidth]{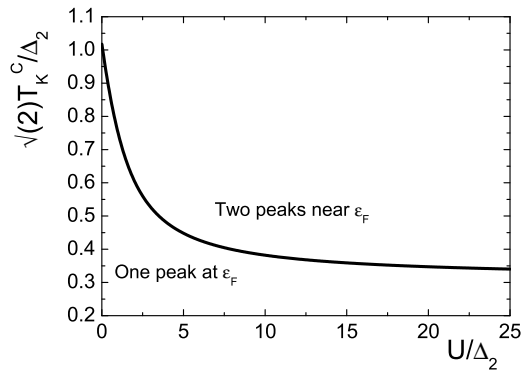} 
\caption{Kondo temperature
as a function of $U$ at the boundary for the appearance of split peaks.}
\label{tkc}
\end{figure}

It has been suggested that in general, the critical condition can be defined
as $\sqrt{2}T_{K}^{c}\simeq \Delta _{2}$, where $T_{K}^{c}$ is the Kondo
temperature at the critical line.\cite{dias} The first member of this
equation as a function of $U$ is represented in Fig. \ref{tkc}, with the
Kondo temperature defined by Eq. (\ref{tk}). Within a factor 3 
we find that this condition is correct, which is not bad in view of
the exponential dependence expected for $T_{K}$ on the parameters near the
transition. We note that due to a cancellation of factors, the same 
$T_{K}^{c}$ is obtained if the roots $R_{\sigma }$ and $L_{\sigma }$ are
dropped in the SBMFA. Therefore, this result seems to be robust. 

\subsection{Kondo temperature and universal behavior}

For constant $\Delta (\omega )$ in the symmetric Anderson model, the Kondo
temperature is given by $T_{K}\simeq D\exp [-\pi U/4\Delta ]$, where $D$ is
the band width, and the properties of the system, depend on $T_{K}$ rather
than $U$ and $\Delta $ individually. Thus, the behavior is universal in the
sense that different curves can be mapped into one with appropriate scaling.
In the present case, when $\Delta _{2}\gg T_{K}$, $\Delta (\omega )$ is flat
on the scale of $T_{K}$, and one expects the same behavior with $\Delta $
replaced by $\Delta (0)=\lambda ^{2}/\Delta _{2}$. Therefore, in this
region, when in general there are no split peaks, one expects $T_{K}\simeq
D\exp [-\pi U\Delta _{2}/(2\lambda )^{2}]$. However, for $\Delta
_{2}\rightarrow 0$, one has a different behavior, given by Eq. (\ref{tkd0}).

\begin{figure}[tbh]
\includegraphics[width=.9\linewidth]{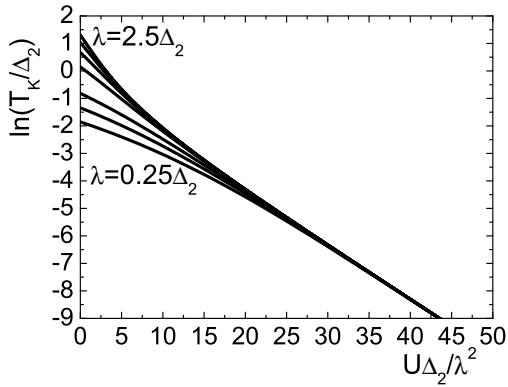} 
\caption{Natural logarithm of the Kondo temperature
as a function of the ratio $U \Delta_2 / \lambda^2$ for several values of 
$\lambda/\Delta_2 $. From bottom to top 0.25, 0.35, 0.5, 1, 1.5, 2, and
2.5}
\label{tkf}
\end{figure}

In Fig. \ref{tkf} we show $\ln (T_{K}/\Delta _{2})$ as a function of the
ratio $U\Delta _{2}/\lambda ^{2}$ for several values of $\lambda $,
calculated with the SBMFA. As expected, for large values of this ratio, for
which there is only one peak near the Fermi energy, all curves merge into
one straight line, indicating a universal behavior, and an exponential
dependence of $T_{K}$ on this ratio. However, as $U\Delta _{2}/\lambda ^{2}$
decreases, already for rather large values ($\sim 10$) for which no split
peaks are expected according to Fig. \ref{lcf}, the values of $T_{K}$ for
different $\lambda $ start to deviate between them and from the exponential
behavior. The decay is faster (slower) than exponential for the larger
(smaller) values of $\lambda $ considered.

\subsection{Comparison of the spectral density obtained by different methods}

\begin{figure}[tbh]
\includegraphics[width=.9\linewidth]{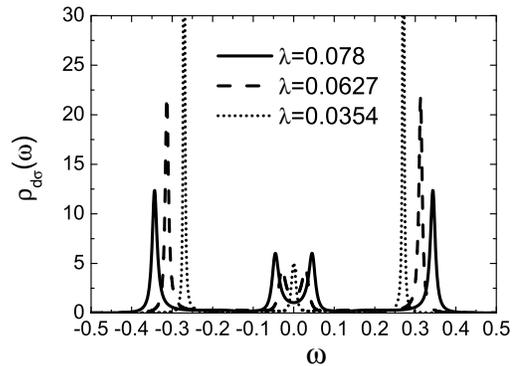} 
\caption{Spectral density of states  for $U=0.5$,  $\Delta_2=0.02$ 
and several values of $\lambda$ calculated with PTU.}
\label{pert}
\end{figure}

In Fig. \ref{pert} we show the spectral density calculated by PTU for the
same parameters as those used in recent NRG calculations.\cite{dias} The
scale is chosen to display the high intensity of the side peaks. 

In general, it is difficult to determine the range of validity of the 
PTU in terms of
the parameters $U$,  $\lambda $ and $\Delta _{2}$. For $\Delta _{2} \gg T_K$
(well inside the region without split peaks), comparison with results of quantum
Monte Carlo \cite{silver} and other calculations \cite{dots} suggest that PTU
is quantitatively valid for $U < 8 \Delta _{0}$, where 
$\Delta _{0}=\lambda^2 / \Delta _{2}$. This condition suggests that PTU is near 
its limit of validity  
for the smallest value $\lambda=0.0354 $ considered in Ref. \onlinecite{dias}.
When $\Delta _{2} < T_K$, one might expect that the above condition should
be replaced by $U < 8 \Delta _{av}$, where $\Delta _{av} < \Delta _{0}$ is 
$\Delta (\omega)$ [see Eq. (\ref{dw}) ] averaged over a range of energies of
the order of $T_K$. In any case, comparison with the results of the SBMFA
discussed below for the other two values of $\lambda $ used, indicates that the
results of the PTU are reliable for these sets of parameters. The comparison
with exact results for $\Delta _{2} \ll \lambda$ discussed in subsection B
also supports the validity of the PTU at zero temperature.

For $\lambda=0.0354$, $\rho _{d\sigma }(\omega )$ reaches 90 in the
arbitrary units chosen in Ref. \onlinecite{dias}. These narrow peaks 
near $-U/2$ and $U/2$ are
probably lost by the NRG approach, which shows only broad features there.
However, these peaks might be experimentally accessible and are therefore
relevant.

\begin{figure}[tbh]
\includegraphics[width=.9\linewidth]{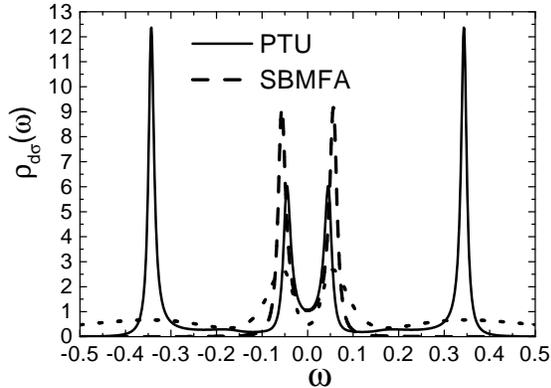} 
\caption{Spectral density of states near the Fermi energy for 
$U=0.5$,  $\Delta_2=0.02$ and $\lambda=0.078$. 
Full line: PTU, dashed line: SBMFA, dotted line: result for
$\Delta_2=0$ broadened as in NRG calculations (see text)}
\label{comp1}
\end{figure}

The density for the larger value of $\lambda $ is shown in Fig. \ref{comp1}
together with the corresponding SBMFA result. The side peaks are lost in the
latter approximation. Both results agree qualitatively. We believe that as
discussed in subsection D, the use of the roots $R_{\sigma }$ and $L_{\sigma
} $[see Eqs. (\ref{r})] in the SBMFA leads to renormalization factor 
$z=(\tilde{\lambda}/\lambda )^{2}$ which is larger than the correct one for
large values of $U$. This is probably the reason of the discrepancy. In any
case both approaches lead to peaks around the Fermi energy which are
three or four times larger than those reported in Ref. \onlinecite{dias}. This again
is likely due to lack of resolution of the NRG results for energies away from
the Fermi energy, as argued below.
Note that reducing $\Delta_2$ to zero, the exact solution [see Eqs. (\ref{r})] gives 
delta functions at $\omega= \pm 0.053$ and $\omega= \pm 0.347$, with weights  
$A=0.221$, and $1/2-A=0.279$ respectively. This result is quite consistent with the
PTU one shown in Fig. \ref{comp1}.
For example, a lorentzian fit of a side peak obtained with PTU gives
a center at $\pm 0.343$, an area 0.284, and a total width 0.0147
(smaller but of the order of $\Delta_2$).
Furthermore, taking these four delta functions artificially broadened
as it is usual in NRG calculations (using Eq. (10) of Ref. \onlinecite{bulla} with
$b=0.6$) and convoluting it with a Lorentzian of total width $\Delta_2$ we obtain the result
displayed by a dotted line in in Fig. \ref{comp1}, which is very similar to that shown in 
Fig. 2 (c) of Ref. \onlinecite{dias}, with central split peaks of intensity near 2
and only very broad bumps replacing the side peaks near $\omega= \pm 0.35$. 
These arguments indicate that the NRG, at least in its usual form, 
is inadequate to describe the spectral 
density in the region of parameters in which split peaks are present.

\begin{figure}[tbh]
\includegraphics[width=.9\linewidth]{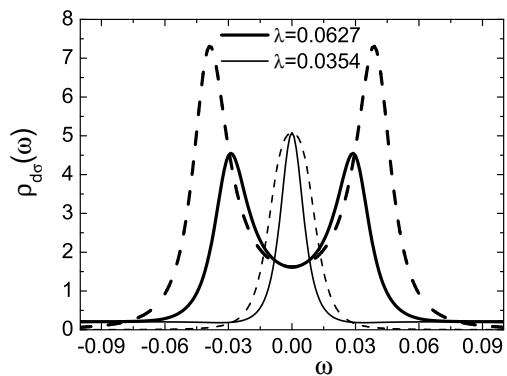} 
\caption{Spectral density of states near the Fermi energy for 
$U=0.5$,  $\Delta_2=0.02$ and two values of $\lambda$. 
\label{comp2}
Full line: PTU, dashed line: SBMFA.}
\end{figure}

The results for the spectral density within PTU and SBMFA near the Fermi
energy for the smaller two values of $\lambda $ are shown in Fig. \ref{comp2}. 
For the same reason discussed above, we believe that the splitting
and intensity of the peaks for $\lambda =0.0627$ within the SBMFA are
exaggerated. As before, the PTU results predict narrower peaks near the
Fermi energy than the NRG results. For the smallest value $\lambda =0.0354$,
we believe that both, the PTU and the SBMFA including the roots $R_{\sigma }$
and $L_{\sigma }$ exaggerate the width of the only peak near the Fermi
energy and the NRG is expected to provide more reliable results. In comparison with
the latter, the SBMFA without the roots seems to predict a too narrow Kondo
peak. However, even in this case, the NRG is not able to describe properly the
narrow side peaks. We expect that this defficency persists as long as $\Delta_2 <U/2$.

\subsection{Effects of temperature}

\begin{figure}[tbh]
\includegraphics[width=.9\linewidth]{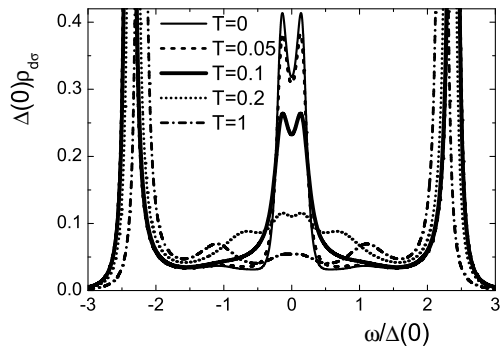} 
\caption{Spectral density of states near the Fermi level 
for $\Delta_2=0.2 \Delta(0)$, $\lambda= \sqrt{\Delta_2 \Delta(0)}$, 
$U=4$ and several values of the temperature $T/\Delta(0)$.}
\label{t}
\end{figure}

In Fig. \ref{t} we show the evolution of the spectral density with
temperature in a region in which there are split peaks, calculated with the
PTU. Similarly to the case of constant $\Delta (\omega )$, the peaks near
the Fermi energy decrease in a temperature scale of the order of the half
width at half maximum of the structure. In contrast to that case however,
some structure remains even at the highest temperatures. This can be
understood qualitatively from the behavior observed in the exactly solvable
limit $\Delta _{2}=0$. 

In addition, the evolution of the side peaks with
temperature is more marked than in the case of constant $\Delta (\omega )$,
probably due to the fact that the broadening effects of the temperature are
more noticeable when the peaks are narrower.

At temperatures higher than $T=\Delta(0)$, there is little variation in 
the spectral density. The result for $T=5 \Delta(0)$ (not shown) coincides
with that for $T=\Delta(0)$ within the width of the line in the figure.
It is curious
that some structure appears near $\omega /\Delta(0)= \pm 1.2$ at high
temperatures. 
Its origin is unclear to us.

\section{Summary and discussion}

We have studied an impurity Anderson model in the symmetric case, in which
the hybridization function $\Delta (\omega )$ has a Lorentzian shape of
width $\Delta _{2}$. The model can be regarded as a simplified version of
that proposed to describe the projection of the Kondo effect in quantum
corrals (the quantum mirage),\cite{rev} and is relevant for systems of
double QD's in which one can be regarded as 
non-interacting and connected to metallic leads
(or other system of non-interacting electrons) and the other is interacting
and connected as a side dot to the non-interacting one.\cite{dias}

The characteristic low-energy scale or Kondo temperature $T_{K}$ of the
model, changes from the known exponential dependence $T_{K}\simeq D\exp
[-\pi U/4\Delta (0)]$ in the Coulomb repulsion $U$ for large $U$ and a flat 
$\Delta (\omega )$ (large $\Delta _{2}$), to $T_{K}=8\lambda ^{2}/U$  for
large $U$ and $\Delta _{2}\rightarrow 0$, where $\lambda $ is the interdot
hybridization. For small enough $\Delta _{2} \lesssim T_{K}$, the scaling properties 
of the Kondo regime are lost, but by extension we continue to call ``Kondo peak" the structure 
near the Fermi energy in the spectral density.

When $\Delta _{2}$ becomes smaller than $ \sim T_{K}/2$ 
(see Fig. \ref{tkc}) the characteristic Kondo
peak in the spectral density $\rho _{d\sigma }(\omega )$ of the interacting
QD near the Fermi energy splits in two. This splitting was reported before
in the context of the mirage effect \cite{hvar} (but not discussed in detail
there) and in recent calculations using 
the numerical renormalization group (NRG).\cite{dias} 
Our calculations using perturbation theory in $U$ and a
slave-boson mean field approximation (SBMFA) agree qualitatively with these results,
but predict narrower split peaks. Moreover, we find that the side peaks of  
$\rho _{d\sigma }(\omega )$  near the charge transfer energies are much
narrower and higher that in the usual case of a flat $\Delta (\omega )$ and
than the results presented in Ref. \onlinecite{dias}. 

In Section III F, we have provided several arguments that 
indicate that the above mentioned discrepancies are due to
shortcomings of the numerical technique. These can be partially
overcome if density matrix renormalization group is used in combination
with NRG.\cite{zit2} Another way of improving the NRG results which has been 
shown to lead to sharper charge transfer peaks is to calculate the self energy 
as a ratio of two Green's functions 
$\Sigma _{dd\sigma }(z)=U \langle \langle d_{\sigma}d_{\bar{\sigma}}^{\dagger }d_{\bar{\sigma}},d_{\sigma}^{\dagger} \rangle \rangle _{z}/ \langle \langle d_{\sigma},d_{\sigma}^{\dagger} \rangle \rangle _{z}$, 
and replace the result in 
Eq. (\ref{gdd}).\cite{bulla} This has the advantage that the effect of $\Sigma _{sd\sigma }(z)$ 
is taken into account exactly.

The splitting of the peaks can be understood qualitatively from the limit 
$\Delta _{2}\rightarrow 0$, and quantitatively 
in terms of weakly interacting renormalized quasiparticles, as described in
Section III D. 

The SBMFA provides an analytical expression, leading to
a diagram for the region of parameters for which split peaks in $\rho
_{d\sigma }(\omega )$ are expected. This is represented in Fig. \ref{lcf}.
In the region in which split peaks are present, some structure remains 
near the Fermi energy even at temperatures much higher than the Kondo 
temperature $T_K$.

The approximations that we have used satisfy Fermi liquid relations and work
well in the limit of small $\Delta _{2}$. In particular the agreement of
perturbative calculations with the exact results for 
$\Delta _{2}\rightarrow 0$ at zero temperature is surprising.

If the non-interacting resonance (the energy of the non-interacting QD)
is shifted away from the Fermi energy in some energy larger than its width
$\Delta _{2}$,  we expect
a dramatic change in the spectral density at low energies, 
with a very narrow Kondo resonance 
at the Fermi energy, due to the decrease of the non-interacting density
of states at the Fermi level, and the exponential dependence of the Kondo
energy scale with this density. 
This is based on previous calculations using approximations 
\cite{rev} and supported by NRG results.\cite{corna}  
Instead, the side peaks should not be affected substantially.

The presence of narrow side bands in the region in which the Kondo peak
splits in two can in principle be tested experimentally in QD systems. The
spectral density at the interacting dot can be measured in transport
experiments in which another lead is added.\cite{leba,letu} 

\section*{Acknowledgments}

This work was sponsored by PIP 5254 of CONICET and PICT 03-13829 of ANPCyT.
AAA and AML are partially supported by CONICET.

\end{document}